\documentclass{article}

\usepackage{PRIMEarxiv}

\usepackage[utf8]{inputenc} 
\usepackage[T1]{fontenc}    
\usepackage{hyperref}  
\hypersetup{
    colorlinks=true,
    linkcolor=blue,
    filecolor=blue,      
    urlcolor=blue,
    citecolor=blue,
}     
\usepackage{url}            
\usepackage{booktabs}       
\usepackage{amsfonts} 
\usepackage{amsmath}
\usepackage{nccmath}
\usepackage{tabularx}
\usepackage{graphicx}
\usepackage{caption}  
\DeclareCaptionFormat{sanslabel}{#3}    
\usepackage{nicefrac}       
\usepackage{microtype}      
\usepackage{lipsum}
\usepackage{fancyhdr}       
\usepackage{graphicx}       
\graphicspath{{media/}}     

\title{Functional Data-Driven Framework for Fast Forecasting of Electrode Slurry Rheology Simulated by Molecular Dynamics}

\author{ {\hspace{1mm}Marc Duquesnoy}$^{1, 3}$ \\
	\texttt{} \\
	\And
	{\hspace{1mm}Teo Lombardo}$^{1, 2}$ \\
	\texttt{} \\
	\And
	{\hspace{1mm}Fernando A. Caro}$^{1, 2}$ \\
	\texttt{} \\
	\And
	{\hspace{1mm}Florent Haudiquez}$^{1}$ \\
	\texttt{} \\
	\And
	{\hspace{1mm}Alain C. Ngandjong}$^{1, 2}$ \\
	\texttt{} \\
	\And
	{\hspace{1mm}Jiahui Xu}$^{1, 2}$ \\
	\texttt{} \\
	\And
	{\hspace{1mm}Hassan Oularbi}$^{1, 2}$ \\
	\texttt{} \\
	\And
	{\hspace{1mm}Alejandro A. Franco}$^{1, 2, 3, 4, *}$}

\begin{document}
\maketitle

{$^{1}$ \small Laboratoire de Réactivité et Chimie des Solides (LRCS), Université de Picardie Jules Verne UMR CNRS 7314, Hub de l’Energie, 80039 Amiens, France.}

{$^{2}$ \small Réseau sur le Stockage Electrochimique de l’Energie (RS2E), FR CNRS 3459, Hub de l’Energie, 80039 Amiens, France.}

{$^{3}$ \small Alistore-ERI European Research Institute, CNRS FR 3104, Hub de l’Energie, 80039 Amiens, France.}

{$^{4}$ \small Institut Universitaire de France, 103 Boulevard Saint Michel, 75005 Paris, France.}

{$^{*}$ \small Corresponding author : alejandro.franco@u-picardie.fr (Alejandro A. Franco)}

\vspace{2cm}

\begin{abstract}
Computational modeling of the manufacturing process of Lithium-Ion Battery (LIB) composite electrodes based on mechanistic approaches, allows predicting the influence of manufacturing parameters on electrode properties. However, ensuring that the calculated properties match well with experimental data, is typically time and resources consuming In this work, we tackled this issue by proposing a functional data-driven framework combining Functional Principal Component Analysis and K-Nearest Neighbors algorithms. This aims first to recover the early numerical values of a mechanistic electrode manufacturing simulation to predict if the observable being calculated is prone to match or not, \textit{i.e} screening step. In a second step it recovers additional numerical values of the ongoing mechanistic simulation iterations to predict the mechanistic simulation result, \textit{i.e} forecasting step. We demonstrated this approach in context of LIB manufacturing through non-equilibrium molecular dynamics (NEMD) simulations, aiming to capture the rheological behavior of electrode slurries. We discuss in full details our novel methodology and we report that the expected mechanistic simulation results can be obtained 11 times faster with respect to running the complete mechanistic simulation, while being accurate enough from an experimental point of view, with a $F1_{score}$ equals to 0.90, and a $R^2_{score}$ equals to 0.96 for the learnings validation. This paves the way towards a powerful tool to drastically reduce the utilization of computational resources while running mechanistic simulations of battery manufacturing electrodes.
\end{abstract}


\section{Introduction}
\paragraph{} In a modern world where Artificial Intelligence (AI) and Machine Mearning (ML) applications are blooming \cite{AI1, AI2}, one may think that the use of more traditional mechanistic models based on mathematical descriptions of physical processes, is becoming obsolete. However, this is not true since mechanistic models still represent nowadays unavoidable tools to support the analysis of complex systems. In contrast to empirical models (and \textit{per se} ML approaches) which study the real world to elaborate a theory, mechanistic models are based on a theory to predict the real-world behavior. Models of that kind are omnipresent in numerous domains such as medicine \cite{Medicine}, battery manufacturing \cite{Phases, PSO}, nanotechnologies \cite{Nano}, biology \cite{Biomechanics, Biophysics}, or in environmental sciences \cite{Environment}. The significant progress in computational hardware achieved in the last decades boosted the emergence of a massive amount of academic and commercial software \cite{LAMMPS, Simulink}, allowing to create and solve the equations behind mechanistic models describing systems with increasing complexity \cite{Cluster1, Cluster2}. This is the case in a plethora of scales, \textit{e.g.} materials properties from their electronic structure \cite{Scale1} or the dynamics of colliding galaxies from gravitational forces \cite{Scale2} and gas hydrodynamics \cite{Scale3}. Furthermore, the capability of a mechanistic model to \textit{simulate} the system behavior by precisely controlling the hypothesis of work, makes possible to perform on-demand \textit{computer experiments} to evaluate the impact of different assumptions on the results. The latter makes this type of models particularly suitable to support analysis of experimental data in both fundamental and applied research activities. \\
Despite the continuous improvement in the numerical methods used to closely match the experimental context of interest with high-fidelity data, computations are usually time and resources consuming. The overall time and mandatory resources required to recover the complete simulation process determine the \textit{computational cost} of the corresponding simulation, and varies as a function of the specific model utilized. A high computational cost narrows the number of possible parameter sets for testing new conditions \cite{CFD}, restraining the use of mechanistic models in a high-throughput way \cite{Use, Cluster4, Cluster5, Compact}.
 
\paragraph{} However, Machine Learning (ML) techniques can be combined with mechanistic models to reduce their computational cost. ML techniques are very popular nowadays to support researchers to move away from pure trial-and-error approaches in experimental contexts, to facilitate properties calculations and to ease the parameterization of mechanistic models \cite{Kernel, ANN}. For instance,  Wang \textit{et. al.} applied different ML algorithms to analyze molecular dynamics simulations \cite{CGMD}, whereas Adam \textit{et. al.} combined neural networks with physics-based models to improve the accuracy of lithographic process modeling \cite{Lithography}. From an industrial perspective, Maleh \textit{et. al.} studied the use of ML techniques for Internet of Things (IoT) intrusion detection in aerospace cyber-physical systems \cite{IOT}, and Tuncali \textit{et. al.} evaluated cyber-physical systems with ML in the context of autonomous driving systems \cite{Driving}. In contrast to usual data analysis techniques for multivariate datasets, \textit{Functional Data Analysis} (FDA) that processes time series, can also be connected with ML techniques, but for the forecasting or prognostic of time-series variables \cite{TimesSeries2}. Therefore, such a combination can constitute a straightforward solution for dealing with wavering data from mechanistic models as part of the computational cost reduction \cite{TimesSeries1}, for example, workflows dedicated to their parameters optimization.

\begin{figure}[!ht] 
	\captionsetup{format=sanslabel}
    \hbox to\hsize{\hss\includegraphics[width=16cm]{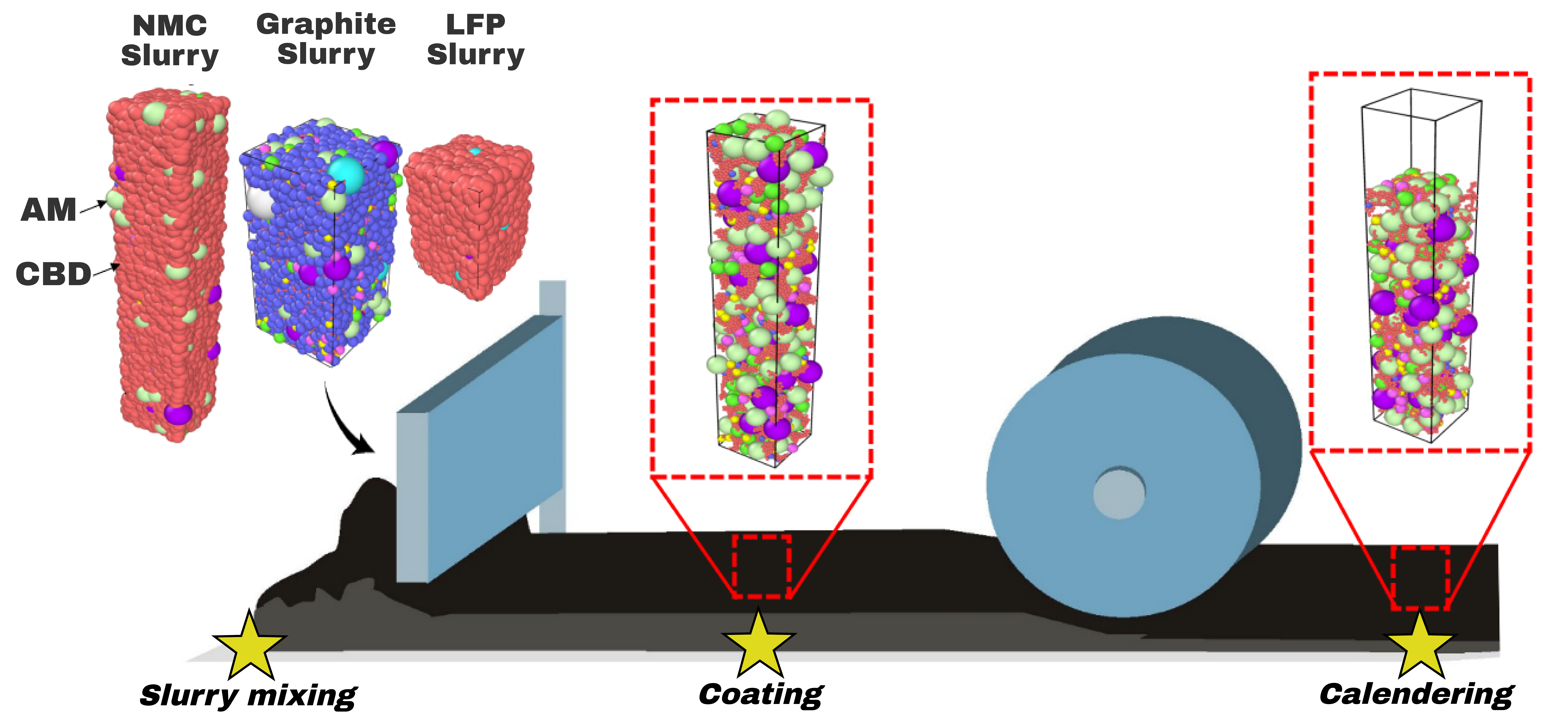}\hss} 
    \caption{ \textbf{Figure 1} : Schematic of a lithium ion battery (LIB) electrode manufacturing process by using mechanistic modeling. Our methodology is illustrated with three different electrode slurry chemistries: Nickel Manganese Cobalt Oxyde (NMC-111), Graphite and Lithium-Iron-Phosphate (LFP).}
\end{figure}

\paragraph{} One of these contexts is process engineering and the particular case of manufacturing process of lithium ion batteries (LIBs) \cite{LIB1}, taken here as an example for the demonstration of our approach. This process encompasses multiple steps and parameters which are interlinked \cite{ManufacturingOpt, Challenge}. Such steps concern the slurry preparation, its coating and drying, the calendering of the resulting electrodes, the cell assembly, the electrolyte infiltration and the formation of the solid electrolyte interphase \cite{Li, Manu1, Manu2, Manu3}. This complex process has been historically simulated using empirical models with parameters fitted to experimental trends or by using mechanistic models based for instance on the Continuum Fluid Dynamics (CFD) approach \cite{2030}. In the recent years, ML algorithms started to emerge in this field, illustrating the powerful capabilities of these techniques to unravel interdependencies between manufacturing parameters and electrode properties \cite{Manu1, Physics2, Physics3}. Furthermore, mechanistic models allowing to predict the electrode microstructure in 3D, as a function of the manufacturing parameters started also to emerge \cite{Phases2, Phases3}. Those 3D mechanistic models account by the explicit spatial location and trajectory of active material (AM) and carbon-binder domain (CBD) phases (Figure 1A) \cite{3D, Phases5}. As an example for the slurry and the coating/drying, these models are supported on the Coarse-Grained Molecular Dynamics (CGMD) approach accounting for attractive and repulsive forces between the considered particles \cite{PSO}. In contrast to such a support, the calendering accounts the Discrete Element Method (DEM) \cite{Phases}, whereas the Lattice Boltzmann Method is used for the electrolyte infiltration modeling. These models have a significant computational cost, but still they need to be run in order to generate high fidelity data (data that generally agrees with the experimental one) for derivation of surrogate models and optimization loops \cite{LBM1}. It requires mechanistic models’ validation by comparison with experimental observables. One of these observables is the slurry viscosity as a function of the applied shear rate. The electrode coating quality strongly depends on this rheological property. In the mechanistic model, once the slurry microstructure is predicted, non-equilibrium molecular dynamics (NEMD) simulations are performed to calculate the \textit{shear-viscosity} curve ($\gamma$-$\eta$ curve). This curve is calculated point by point, \textit{i.e.} a shear-rate is applied through the deformation of the simulated slurry box and a viscosity value is retrieved (see details in Supporting Information). Each calculation must to be run for a time long enough to be confident of having reached a stable $\eta$ value, usually leading to overall high computational costs.

\paragraph{} In this work, we tackled the issue of computational cost reduction of 3D-resolved mechanistic models, through the use of a functional data-driven framework for fast predictions of their simulation results. Briefly, this consists at running the first numerical steps (\textit{i.e.} time frames) of the mechanistic simulation, to retrieve early results, and then bypass the full simulation process by predicting its results. More precisely, the aforementioned framework proposes first a screening step, whose main goal is to identify running simulations of interest from an experimental perspective, \textit{e.g.} expected to give results in agreement with an experimental target value (viscosity for a given shear rate in the application case used here as an illustration). Second, it proposes a forecasting step to quickly predict their results, considering only the previously filtered simulations. Both steps couple two algorithms: one based on \textit{Functional Principal Component Analysis} (FPCA) achieving a compression of time series in a low dimensional space (\textit{i.e.} dimensionality reduction), and another one based on \textit{K-Nearest-Neighbors} (KNN) performing the predictive task. We applied this framework for \textit{in silico} slurry modeling based on NEMD (see details in the Supporting Information) after accumulating simulations over three different LIB electrode chemistries \cite{NEMD1}. We tracked the evolution of the calculated viscosity values ($\eta$) along the simulation process to determine the time series. Despite the illustration here for LIB electrode slurries, the proposed framework can be applied to other fields where mechanistic models generate time series and can trigger on significant computational cost diminution for faster optimization process of the mechanistic model parameters to match experimental observables.

\section{Time series processing}

\subsection{Functional Data Analysis}
\paragraph{} NEMD calculations of electrode slurry viscosities generate discretized values at each numerical step (\textit{i.e.} time frame) along the calculation process. Those steps are independent from the allocated resources used, and produce a list of numerical values to constitute a curve as a function of time. Such a curve is the descriptor of one simulation, and an appropriate data treatment of this requires to fix the number of time frames for the definition of the corresponding \textit{functional} space $\mathbf{I}$ \cite{Space}. The functional form of the cure is written \\

\begin{ceqn}
\begin{align}
\tag{1}
	\begin{split}
		\mathbf{X} = \{X(t), \hspace{0.1cm} t \in \mathbf{I}\} \quad \mathbf{I} \subseteq \rm I\!R^+
	\end{split}
\end{align}
\end{ceqn}

\noindent as opposed to a collection of functional variables associated to one simulation, written as \\

\begin{ceqn}
\begin{align}
\tag{2}
	\begin{split}
		\mathbf{X} = \{X^{(1)}(t),..., X^{(p)}(t), \hspace{0.1cm} t \in \mathbf{I}\} \quad \mathbf{I} \subseteq \rm{I\!R^+}, p \in \rm{I\!N}
	\end{split}
\end{align}
\end{ceqn}

\paragraph{} \textit{Functional Data Analysis} (FDA) properly carries out those types of variables by defining highly regular data over the space $\mathbf{I}$ through smoothing techniques. FDA as shown to be very relevant approach in numerous contexts, as for instance in clinical studies \cite{Kinetics}, sport performance analysis \cite{Sport}, or materials discovery \cite{Materials, Materials2}. Within this study, we applied a \textit{Functional Principal Component Analysis} (FPCA) on time series as a tool to achieve their data compression, \textit{i.e.} dimensionality reduction, for further predictive tasks. In particular, FPCA balances between the deep of the compression and the resulting variability, to obtain a meaningful low dimensional representation of time series \cite{FPCA, MFPCA}. Such a data processing is already well known within the usual \textit{Principal Component Analysis} (PCA) \cite{PCA}, but here it is extended to the functional form of time series. 

\paragraph{} FPCA requires a smoothing technique to reconstruct a set of discrete values from time series (Eq. 1), over the functional space. The latter space is characterized by a set of basis functions, that enables a basic decomposition of time series in a finite dimensional space \cite{Finite}. Let's consider time series like written in (1), and a set of basic functions $\mathbf{\phi} = \{\mathbf{\phi_1}, \mathbf{\phi_2}, ..., \mathbf{\phi_p}\}$ all defined on $\mathbf{I}$ with $p$ in $\rm{ I\!N}$. $\mathbf{X}$ is given by \\

\begin{ceqn}
\begin{align}
\tag{3}
	\begin{split}
		\mathbf{X} = X(t) = \sum_{i=1}^p \mathbf{c_i} \times \mathbf{\phi_i}(t) \quad (\mathbf{c_i}, t) \in \rm{I\!R} \times \mathbf{I}
	\end{split}
\end{align}
\end{ceqn}

\noindent where the set $\textbf{c} = \{\mathbf{c_1}, \mathbf{c_2}, ...,\mathbf{c_p}\}$ is the vector of basis coefficients, which describes $\mathbf{X}$ by a finite number of values. 

\paragraph{} The most common solution for the basis decomposition is to use a \textit{B-Spline} family \cite{BSpline2}. The associated methodology consists at interpolating time series with a family of polynomial functions (basic functions), which are sufficient differentiable in discrete knots. Considering $\mathbf{I} = [\mathbf{I_0}, \mathbf{I_1}]$ with $\mathbf{I_0} = \mathbf{t_0} < \mathbf{t_1} < ... < \mathbf{t_{p}} = \mathbf{I_1}$, each polynomial is defined in sub-intervals of $ \mathbf{I}$, expressed by $(\mathbf{t_i})_{(i \leq p)}$ values. The latter polynomials are positive on at the most $p$ sub-intervals, and their degree is equal to 3 in this study.

\paragraph{} Based on the expression in (3), FPCA can estimate $M$ eigenfunctions $(\mathbf{\psi_i})_{(i \leq M)}$ associated to $M$ eigenvalues $(\mathbf{\nu_i})_{(i \leq M)}$, and finally $M$ scores $(\mathbf{\rho_i})_{(i \leq M)}$ representing the projection of time series along new axis, also called functional principal components. In particular, $M$ is very limited compared to $p$, since the FPCA must keep as much variability (variance) as possible from the initial time series in a short amount functional principal components, and illustrates the concept of dimensionality reduction.\\
Regarding the usual PCA \cite{PCA, Trace}, the covariance matrix denoting correlations between random variables is replaced by a covariance operator $\chi$, due to the functional form of time series. It allows the assessment of the eigenvalues associated to the eigenfunctions (extension of the Hilbert-Schmidt theorem \cite{Hilbert}) of $\chi$, according to its approximation based on the above basis decomposition. From the linear algebra, the resulting FPCA is expressed by $\mathbf{X}$ following the Karhunen-Loève decomposition \cite{Karhunen} \\

\begin{ceqn}
\begin{align}
\tag{4}
	\begin{split}
		\mathbf{X} = \sum_{m=1}^{M} \mathbf{\rho_{i,m}} \times \mathbf{\psi_m} 
	\end{split}
\end{align}
\end{ceqn}

\paragraph{} The extension of the \textit{Singular Values Decomposition} (SVD) \cite{PCA} to functional forms for time series explicits a straightforward relationship between $\mathbf{\psi_m}$ and $\mathbf{\nu_m}$ for all $m$, being \\

\begin{ceqn}
\begin{align}
\tag{5}
	\begin{split}
		\mathbf{\chi} \times \mathbf{\psi_m} = \mathbf{\nu_m} \times \mathbf{\psi_m} \quad 
	\end{split}
\end{align}
\end{ceqn}

\paragraph{} In our study, we applied the same basis decomposition, \textit{i.e.} basis family, for the smoothing of each time series $(X_j)_{j \leq N}$ and eigenfunctions $\psi_j(t) = \sum_{i=1}^p \mathbf{b_{i, j}} \times \mathbf{\phi_i}(t) \hspace{0.2cm} j \leq M$. As a result, (3) becomes matrix-wise \\

\begin{ceqn}
\begin{align}
\tag{6}
	\begin{split}
		X_j(t) = \sum_{i=1}^p \mathbf{c^{(j)}_i} \times \mathbf{\phi_i}(t) \Leftrightarrow \mathbf{X} = \mathbf{C} \Phi \quad (\mathbf{c^{(j)}_i}, t, j)\in \rm{I\!R} \times I \times \rm{I\!N}
	\end{split}
\end{align}
\end{ceqn}

\noindent with $\mathbf{C}$ and $\Phi$, the matrix of basis coefficients and the vector of basis functions respectively.

\paragraph{} Accordingly to previous expressions, the initial problem (5) can be shaped as the following product of matrices to link any couple of eigenfunction and eigenvalue \cite{FPCA} \\

\begin{ceqn}
\begin{align}
\tag{7}
	\begin{split}
		\frac{1}{N} \Phi^T \mathbf{C}^T \mathbf{C} W \mathbf{b} = \mathbf{\nu} \Phi \mathbf{b}
	\end{split}
\end{align}
\end{ceqn}

\noindent where $W$ is included in $\rm{I\!R}^{p\times p}$,  containing the pairwise scalar products of basis functions $([<\mathbf{\phi_i}, \mathbf{\phi_j}>_2])_{(p)}$. $\mathbf{b}$ is the vector of coefficients from the basis decomposition of $\psi$.\\

\paragraph{} We denote $\mathbf{a} = W^{1/2}\mathbf{b}$ to simplify the equation above, allowing to write \\

\begin{ceqn}
\begin{align}
\tag{8}
	\begin{split}
		\frac{1}{N} W^{1/2} \mathbf{C}^T \mathbf{C} W^{1/2} \mathbf{a} = \mathbf{\nu} \mathbf{a}
	\end{split}
\end{align}
\end{ceqn}

\paragraph{} Such a final equation relates an eigenproblem that is common for a SVD in an usual PCA, where the eigenvalues are associated to the matrix $\frac{1}{N} W^{1/2} \mathbf{C}^T \mathbf{C} W^{1/2}$ \cite{PCA, Trace}. Therefore, we obtain $\mathbf{a}$, then calculate $\mathbf{b}$ to find any $(\mathbf{\psi_i})_{(i \leq M)}$, and lastly $(\mathbf{\rho_i})_{(i \leq M)}$.

\paragraph{} At the end of the FPCA, the resulting scores are sufficient to describe each time series in a discret manner, to considerably reduce the number of values. This benefits to further supervised ML implementations of our framework for predictive tasks. The main advantages concern the possibility to decrease the training time due to a narrowed number of possible variables combinations, but also to avoid an overfitting during the training step \cite{Overfitting1, Overfitting2}. The latter mentioned fitting usually happens when the feature space becomes increasingly sparse with a high number of parameters (\textit{i.e.} inputs), because ML models can be very sensitive to a short set of parameters \cite{Sensitive}. Therefore, the application of dimensionality reduction for time series helps in that sense, to represent a meaningful data processing for defining inputs of ML models. 

\subsubsection{Physical relevance from an experimental perspective}
\paragraph{} In our framework, we coupled such a FPCA with a ML algorithm performing two different predictive tasks, being the two pillars of our study. The first one concerns an early identification of running simulations of physical relevance from an experimental perspective, commonly called \textit{supervised classification} in the ML terminology. In other words, it means that the framework achieves a screening of mechanistic simulations in the very first numerical steps, by predicting if such a simulation will be ending with relevant results for further experimental comparisons. In our context of slurry modeling on NEMD, there is a physical relevance when the results associated to a simulation (\textit{i.e.} slurry viscosity) are similar to experimental target values. At the end, the prediction provided by the framework concludes on the relevance of each running simulation to let it continue. It represents an automatized "go" or "no go"decision through the analysis of how a simulation behaves.

\paragraph{} In terms of data processing, each simulation is labeled by a binary value representing the output of the supervised classification. This output depends on the final behavior of the simulation like Figure 2A displays. Indeed, we used the average of the numerical values resulting from the last 1000 time frames to consider the result of a simulation. In contrast to a simulation with a high result that suggests to stop its process, a low result recommends to let the simulation running, because of the agreement of the value with the experimental one. As a consequence, we defined a threshold to assess the label, distinguishing high and low results. Figure 2B displays the distribution of the average slurry viscosity, \textit{i.e.} simulations result, that are available in our study. In accordance with an experimental perspectives and the same Figure 2B, we empirically set the threshold at 50 $Pa.s$.

\newpage
\begin{figure}[!ht] 
	\captionsetup{format=sanslabel}
    \hbox to\hsize{\hss\includegraphics[width=17cm]{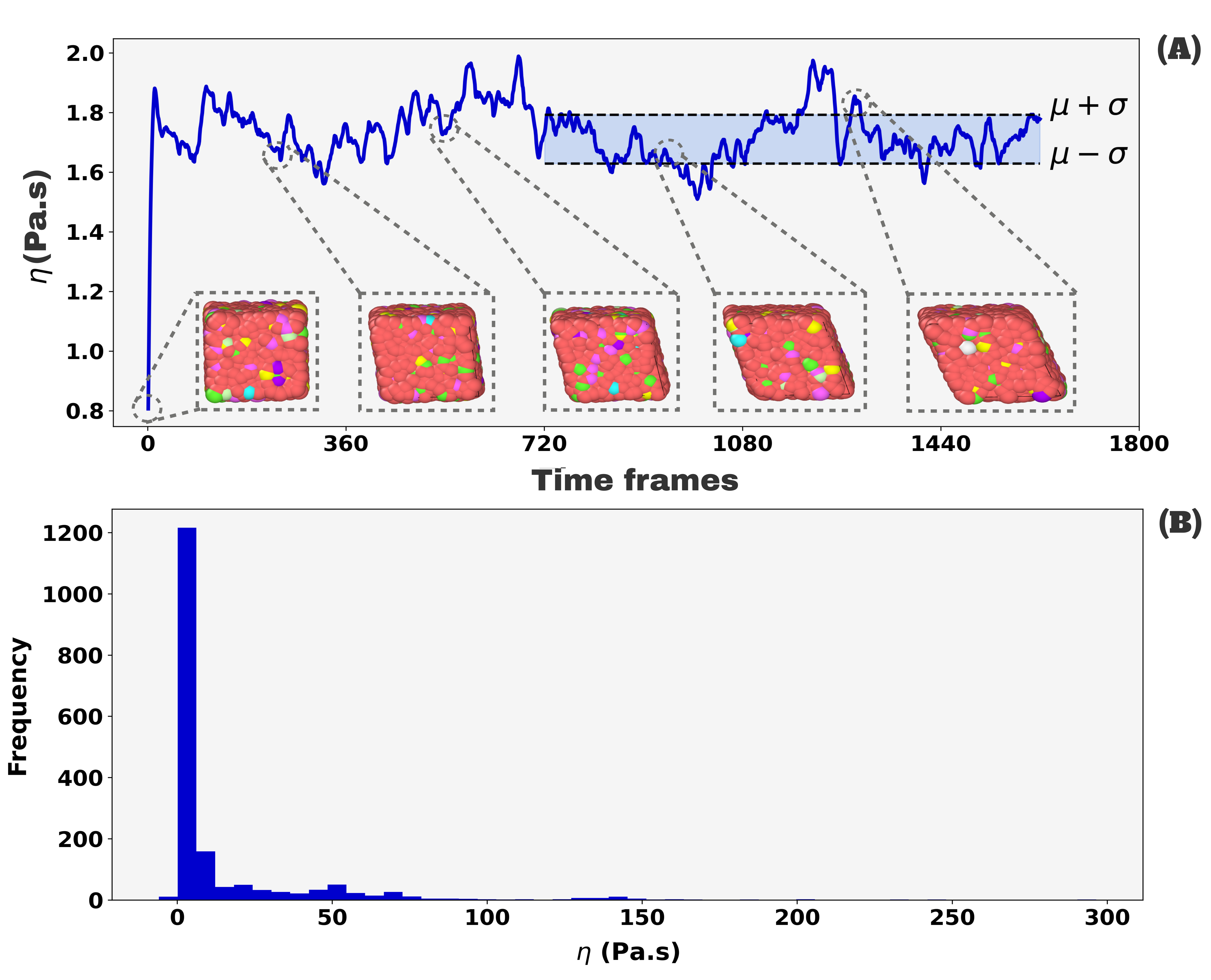}\hss} 
    \caption{ \textbf{Figure 2} : \textit{(A)} : Slurry mesostructure evolution and associated viscosity ($\eta$) during the non-equilibrium molecular dynamics (NEMD) simulation utilized to assess the slurry viscosity at a given shear-rate ($\gamma$) \cite{PSO}. The last 1000 viscosity values from a single simulation are used as a vector, to calculate an average value ($\mu$) and a standard deviation ($\sigma$). These latter values enable to color-code the slurry viscosity evolution within a box, where the two limits are the average plus minor the standard deviation ($\mu$ +/- $\sigma$). \textit{(B)} : Empirical distribution of the average slurry viscosity value ($\eta$) associated to the mechanistic simulations carried out in this study. This show the heterogeneous distribution of represented values.}
\end{figure}

\subsubsection{K-Nearest-Neighbors}
\paragraph{} This screening step (classification) within the aforementioned framework is done using a  \textit{K-Neareast-Neighbors} (KNN) classification \cite{KNN3}. One basis of the algorithm is the memorization of distances between data from the features space. The latter of which is defined by the functional principal components from the FPCA, enhancing calculations of sparse distances between time series (inputs) to make it a meaningful choice for our study \cite{KNN-Dimension}. The KNN algorithm lies on the choice of $k$ neighbors and a distance metric to evaluate pairwise distances between input data, to then attribute a prediction (output) for a new unseen input data, based on the outputs of its $k$ nearest neighbors \cite{KNN-Classification}. Moreover, a right choice of $k$ is crucial to take into account a suitable number of neighbors when predicting the output. This affects the predictive capabilities of the KNN. In this direction, the algorithm is straightforward and can be summarized as it follows 

\begin{enumerate}
    \item[$\textbf{i}$] Choose the number of neighbors $k$;
    \item[$\textbf{ii}$] Calculate the distance between data considering a specific metric;
    \item[$\textbf{iii}$] Get the $k$ nearest neighbors for a new data we want to predict;
    \item[$\textbf{iv}$] Assign the associated output by a majority vote (classification), or by an average value (regression), for the outputs of its $k$ nearest neighbors.
\end{enumerate}

\noindent Section S2 in the Supporting Information provides a graphical interpretation of the KNN algorithm for the case of classification and regression.\\

\begin{figure}[!ht] 
	\captionsetup{format=sanslabel}
    \hbox to\hsize{\hss\includegraphics[width=18cm]{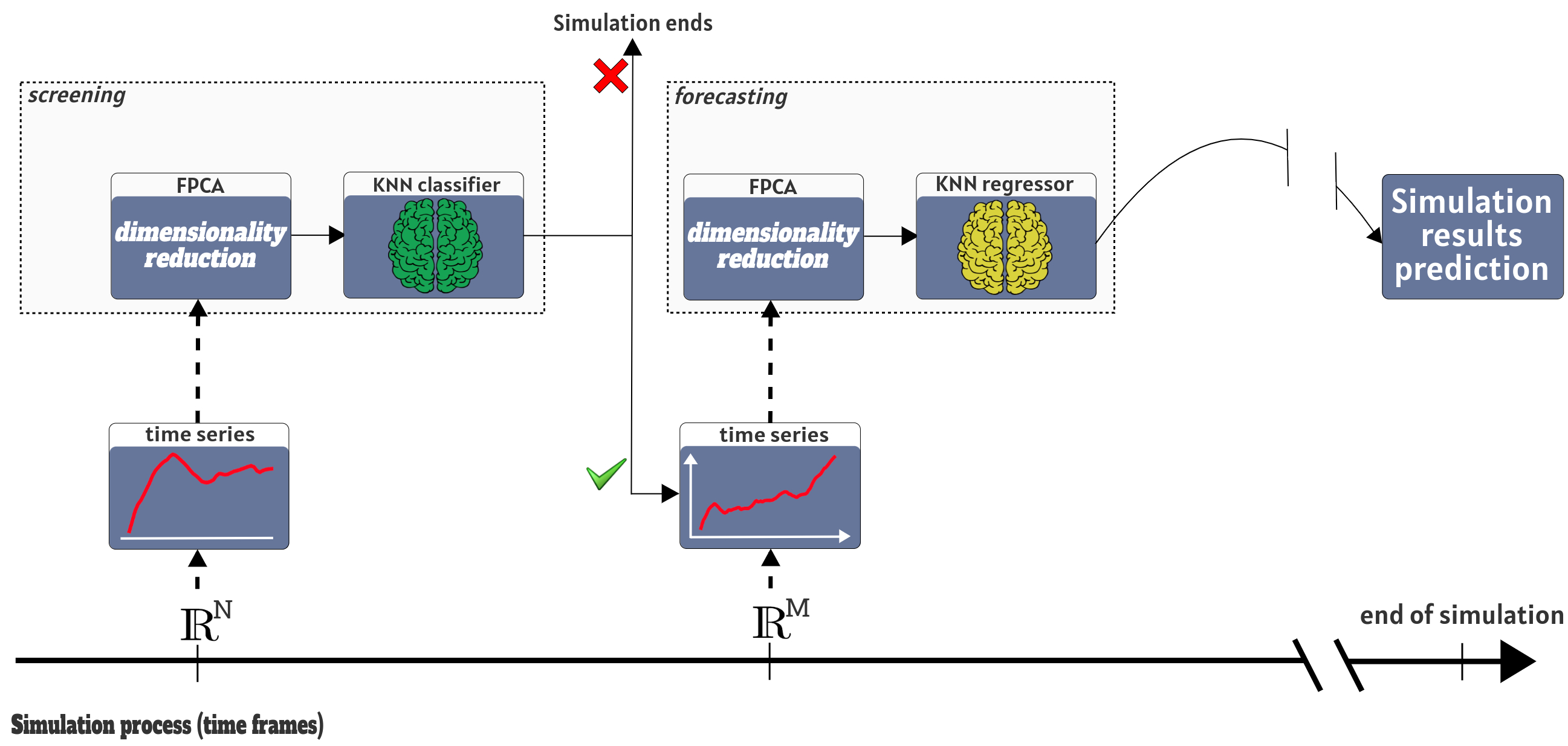}\hss} 
    \caption{ \textbf{Figure 3} : Schematic representation of the functional framework with the different steps, to fast predict the mechanistic simulation results. $\rm{I\!R}^{N}$ and $\rm{I\!R}^{M}$ characterize the definition spaces of time series, with $N$ and $M$ ($N \leq M$) the number of numerical values already generated by the simulation. Such a framework illustrates how a running simulation is handled along its process within early numerical steps, to predict its results and narrow its computational cost.}
\end{figure}

\paragraph{} From the perspectives of the framework, the chosen number of time frames to determine the time series in the screening step must be low, reflecting the efficiency to predict a further physical relevance of a running simulation within the early numerical steps. In the meantime, this enables to stop quickly a running simulation if the latter is predicted as a case without agreement with experimental values. It is particularly relevant for freeing up computational resources and launching another simulation with new operating conditions. This is discussed below by comparing the predictive capabilities of the KNN classification as a function of the number of time frames for time series definition.

\subsection{Fast predictions from mechanistic simulations}
\paragraph{} The second pillar of the aforesaid framework concerns the fast prediction of the results, only for mechanistic simulations that have been previously considered of physical relevance from an experimental perspective. In this context, the results represent the informations we can extract from the last numerical steps of the simulation. As Figure 2A shows, it is the average value ($\mu$) and the standard deviation ($\sigma$) from the vector formed by the numerical values of slurry viscosities within the last 1000 time frames. These results provide not only the final average viscosity of the associated simulation, but also how the latter behaves. To achieve this forecasting step, we also couple a FPCA with a KNN algorithm, but in the context of a \textit{supervised regression} for the latter algorithm with here two outputs, \textit{i.e.} $\mu$-$\sigma$ and $\mu$+$\sigma$.

\paragraph{} Nevertheless, to still tackle the issue of high computational cost of the mechanistic simulation process, we used another number of time frames to define time series for the dimensionality reduction (FPCA). In practice, a simulation is run to be confident of having reaching a stable value to define its final result (Figure 4). This suggests that by recovering the numerical values within a large number of time frames, we may expect the KNN regression to have predictions very closed to the simulation result. In the meantime, this action does not reduce the computational cost since the simulation runs for a large number of time frames. As a consequence, the forecasting step balances between the number of time frames to use for the definition of time series, and the corresponding computational cost to fast predict the simulation result, with high predictive capabilities for the KNN. As we did for the screening step, we compared the predictive capabilities of the KNN regression for different number of time frames to evaluate the best compromise. The results are discussed below.

\begin{figure}[!ht] 
	\captionsetup{format=sanslabel}
    \hbox to\hsize{\hss\includegraphics[width=18cm]{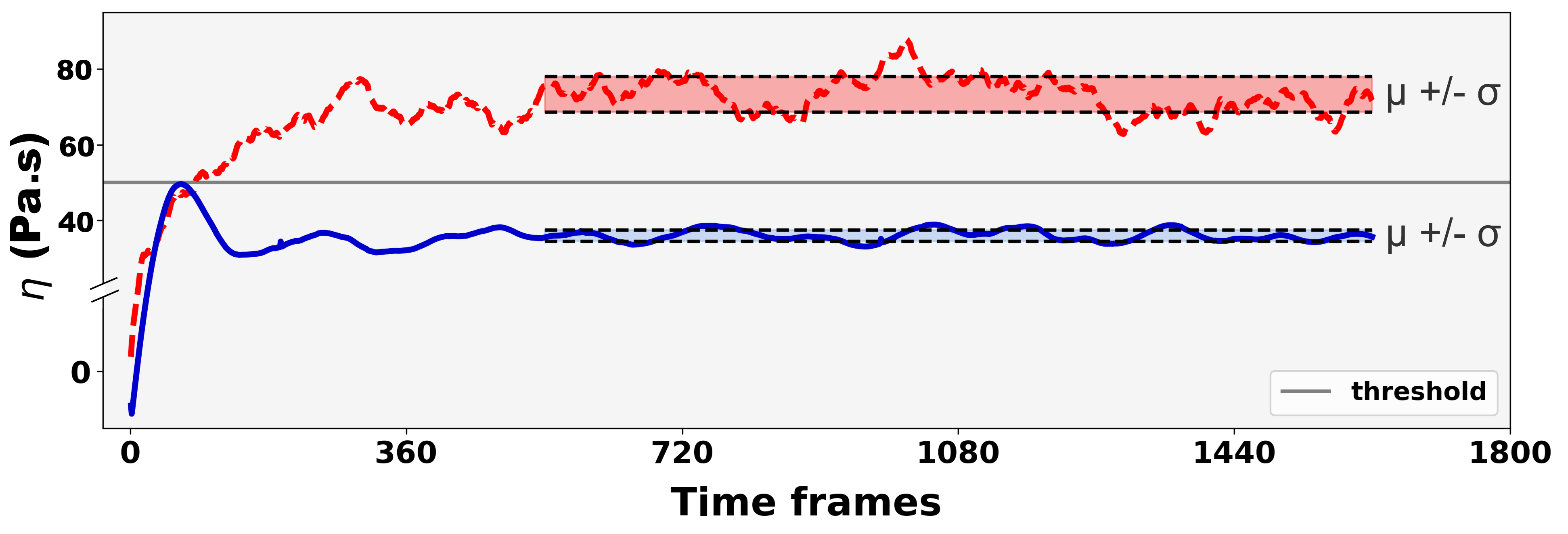}\hss} 
    \caption{ \textbf{Figure 4} : Examples of electrode slurry viscosity calculations with two different behaviors. In contrast to the simulation color-coded in red, the blue one is considered with physical relevance from an experimental perspective. The threshold for the screening step is highlighted by the grey horizontal line at 50 $Pa.s$.}
\end{figure}

\section{Results and Discussion}
\subsection*{Datasets}
\paragraph{} Time series were extracted from NEMD simulations to capture the rheological behavior of LIB slurries (viscosity vs. shear-rate). Simulations were launched using LAMMPS software \cite{LAMMPS}. In total, we executed 2172 simulations coming from the modeling of three different slurries: 1773 from Nickel-Manganese-Cobalt (NMC) slurries, 183 from graphite and 216 from Lithium-Iron-Phosphate (LFP). We notice that the type of chemistry does not determine an input value for the KNN algorithm, since only the corresponding rheological behavior was used for our data processing. Moreover, our code under LAMMPS was executed using the MatriCS HPC platform from Université de Picardie Jules Verne (Amiens, France) \cite{HPC}. In total, the number of time frames to reach stable viscosity values ($\eta$) was between 1450 and 1800, with an average number equals to 1734.\\
In the screening step, all executed simulations complete a functional dataset to train and test our KNN classification. The training dataset contained 80 \% of simulations randomly picked up from the functional dataset, and the testing dataset contained the remaining 20 \%. In the forecasting step, only simulations labeled with physical relevance were used to form another functional dataset to train and test our KNN regression, whose size is equal to 1927 simulations. 

\subsection*{Time frames selection}
\label{Real}
\paragraph{} We chose the number of time frames to set each time series according to Figure 5, which displays the trends for the predictive capabilities, \textit{i.e.} validation metrics, as a function of the number of time frames to set time series for the screening and forecasting step. The same Figure 5 reflects the best compromise between the maximization of the validation metrics and the maximization of the computational efficiency, inversely proportional to the computational cost. In our context, the higher the time to predict the simulation results, the lower the computational efficiency. More precisely, we calculated a numerical mean of the validation metric per number of time frames, by renewing a training/testing dataset 20 times. It is particularly interesting to avoid a calculation of the validation metrics biased by a one single random split of the data into the training and testing datasets. The selected validation metrics are detailed in the methods section below.

\newpage
\begin{figure}[!ht] 
	\captionsetup{format=sanslabel}
    \hbox to\hsize{\hss\includegraphics[width=18cm]{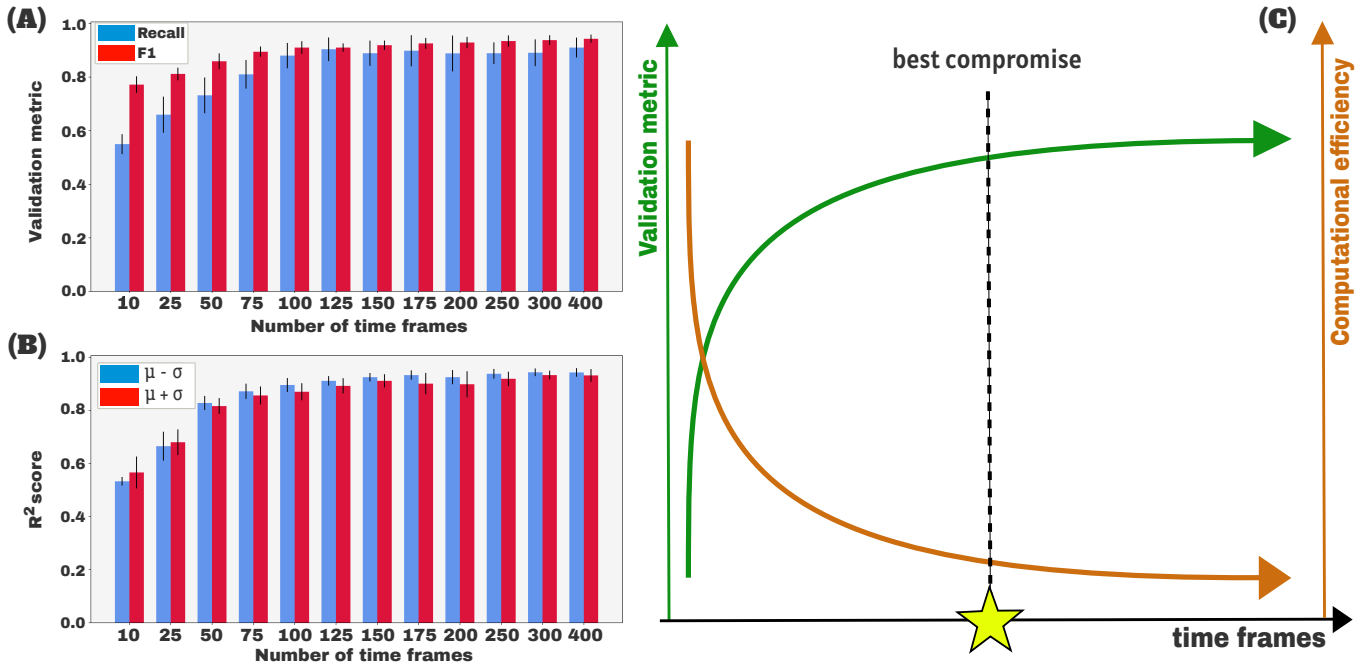}\hss} 
    \caption{ \textbf{Figure 5} : \textit{(A)} : Evolution of the average \textit{Recall} and $F1_{score}$ as a function of the number of time frames. The latter number define the time series needed for the FPCA within the screening step of the framework. The black slight bar represents the deviation around the mean value of the validation metric. \textit{(B)} : Evolution of the average $R^2_{score}$ as a function of the number of time frames. The latter number define the time series needed for the FPCA within the forecasting step of the framework. \textit{(C)} : Illustration of the compromise foreseen, between the selected number of time frames and the corresponding computational efficiency to calculate the simulation results. The green trend illustrates the results from \textit{(A)} and \textit{(B)}. At the end, the best compromise is located for a number of time frames from where the validation metrics are not significantly increasing in order to narrow the computational cost.}
\end{figure}

\paragraph{} We retained a number of time frames equals to 100 for the screening step, whereas the number for the forecasting step was equal to 150. Indeed, Figure 5A and Figure 5B capture increasing trends of validation metrics when incrementing the number of time frames, while starting to stagnate from those two chosen numbers. At the end, the corresponding validation metrics were in average equal to 0.84 and 0.90 for the \textit{Recall}, $F1_{score}$ (KNN classification), and equal to 0.97 and 0.96 for the $R^2_{score}$ of the two respective outputs (KNN regression). Figure 6 displays \textit{regression plots} related to the chosen KNN regression, for the comparisons of predicted values as a function of their corresponding simulation results ($\mu$ +/- $\sigma$). This visualization illustrates how predictions are close with real simulation results.

\paragraph{} In terms of computational cost reduction, the framework enables to use time series defined over 100 and 150 time frames to accurately predict meaningful results from a running simulation. By considering the average number of time frames for all available simulations detailed above, we elaborated a framework efficient to determine the slurry viscosity 11 times faster. In the meantime, the same framework determines the physical relevance of a simulation here 17 times faster. Those results illustrate a drastic reduction of the time required to recover the simulation results, and also to provide quick information on how a given simulation behaves. This is essential to discard running simulations without physical relevance from an experimental perspective, and then avoid the analysis on simulation results that can not be consistent with an experimental comparison. Moreover, the fast prediction of simulation results (slurry viscosity) enables to allocate free available resources on a HPC platform for another simulation to fast study new operating conditions. In our context of slurry modeling, and especially of battery manufacturing, such advantages to fast predict simulation results, facilitate the seek of the best parameter set among extensive possibilities, to trigger further and faster optimization processes of mechanistic model parameters. 

\newpage
\begin{figure}[!ht] 
	\captionsetup{format=sanslabel}
    \hbox to\hsize{\hss\includegraphics[width=16cm]{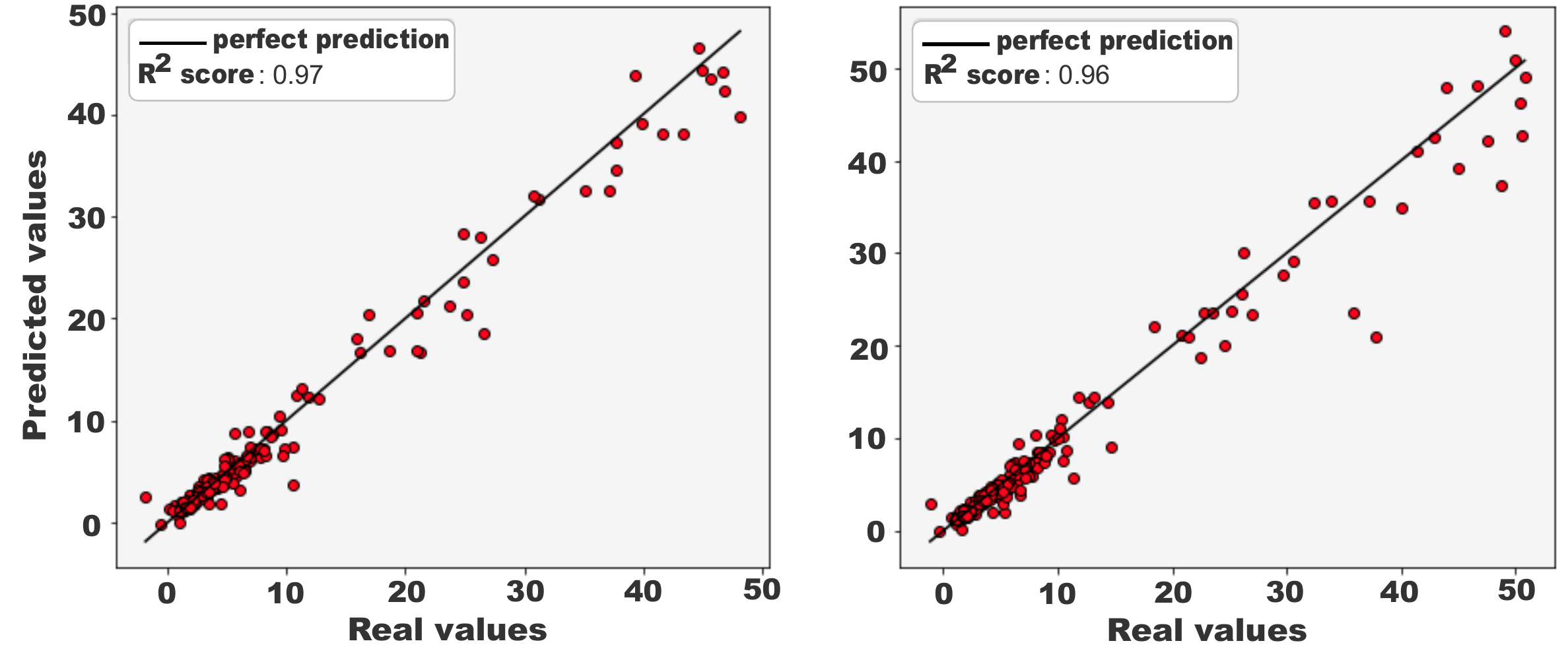}\hss} 
    \caption{ \textbf{Figure 6} : Regression plots to evaluate the predictive capabilities of the KNN regression selected in the data-driven framework, using the testing dataset. The two plots compared the predicted outcomes by the regression as function of the real outcomes from NEMD simulations. The plot of the left corresponds to the output $\mu - \sigma$ whereas the one of the right corresponds to the output $\mu + \sigma$. For each case, the $R^2_{score}$ was rounded with two significant digits.}
\end{figure}

\section{Conclusions}
\paragraph{} In this study, we proposed a functional data-driven framework for fast predictions of mechanistic simulations results, from the very first numerical iterations solving the underlying physical equations. We demonstrated this data-driven approach for the case of the calculations of the viscosities versus applied shear rate for lithium ion battery electrode slurries, with different active material chemistries. Such calculations are supported on Non-Equilibrium Molecular Dynamics and are usually computational expensive. The aforementioned framework quickly determined slurries viscosities within the forecasting step, leading to a computational cost reduction by a factor 11. Besides, such a framework enabled to predict the relevance of a mechanistic simulations from an experimental perspective, here 17 times faster within the screening step. In terms of predictive capabilities, the KNN classification reached a $F1_{score}$ equals to 0.90, whereas the KNN regression achieved a $R^2_{score}$ equals to 0.96. Those validation metrics were considered as reliable from an experimental point to analyze viscosities. In particular, the results in this work presents the capabilities to treat mechanistic simulations as time-series, for the application of FDA techniques. In this way, it serves to support different supervised ML algorithms to accurately analyze the rheology of electrode slurries. In addition, the proposed framework can be transfered to any other field where mechanistic models generate time series, to figure out drastic decrease of corresponding computational costs for quicker parameters optimization processes to design better physics-based models.

\section*{Methods}
\subsection*{Validation metrics}
\label{Metrics} We evaluated the KNN classification and regression by using two different metrics on the testing datasets. The $F1_{score}$  \cite{F1} and especially the corresponding Recall score, were used to validate the KNN classification, whereas we assessed the goodness of fitting for the KNN regression with the $R^2_{score}$ \cite{R2}. The first mentioned metric adjusted the \textit{Precision} and \textit{Recall} (9) to increase the capabilities of a binary classifier to face the issue of wrong predictions. In our context, this is especially relevant with an unbalanced number between mechanistic simulations which will give a correct result and the ones which will not within the dataset. As an example, we considered to minimize the error from the KNN classification to forecast a running simulation with physical relevance, though its real output is without interest \\

\begin{ceqn}
\begin{align}
\tag{9}
    Precision = \frac{TP}{TP + FP} \hspace{2cm} Recall = \frac{TP}{TP + FN}
\end{align}
\end{ceqn}

\noindent where the Recall allows the interest to minimize $FN$, as expressed in the statement above \cite{PrecisionRecall}.\\

\begin{table}[h]
\centering
\small
\setlength\tabcolsep{3pt}
\setlength\extrarowheight{2pt}
\captionsetup{format=sanslabel}

\begin{tabularx}{\textwidth}{ 
  >{\raggedright\arraybackslash}X 
  >{\raggedright\arraybackslash}X }
\midrule
\textsc{KNN classification' instance}   & \textsc{Interpretation}       \\ \midrule
True Positive (TP)      &  Irrelevant simulations predicted as irrelevant. \\
True Negative (TN)      &  Relevant simulations predicted as relevant. \\
False Positive (FP)      & Relevant simulations predicted as irrelevant. \\
False Negative (FN)      &  Irrelevant simulations predicted as relevant.\\
    \midrule          
\end{tabularx}
\caption{\textbf{Table 1} : Summary of the possible instances, \textit{i.e.} predictions, in the case of the KNN classification, for the screening of mechanistic simulations. For the sake of simplicity, we replaced the \textit{"with physical relevance from an experimental perspective"} by \textit{"relevant"}, and its opposite case by \textit{"irrelevant"}.}
\end{table}

\paragraph{} The second mentioned metric ($R^2_{score}$) calculated a to assess how close are predicted simulation results from their real results. The higher the score, the lower the discrepancy between the prediction from the real value is. Such a score is calculated through the equation as it follows \\

\begin{ceqn}
\begin{align}
\tag{10}
    R^2_{score} = 1 - \sum_{i=1}^n (\mathbf{y_i} - \mathbf{\Tilde{y_i}})^2 / \sum_{i=1}^n (\mathbf{y_i} - \mathbf{\Bar{y}})^2
\end{align}
\end{ceqn}

\subsection*{Computational details}
\label{Hyperparameter tuning and final framework}
\paragraph{} The functional framework applies a FPCA on time series for the screening and forecasting step, using a B-spline decomposition with a number of knots equals to 60 \% of the time series length. FPCAs provided a number of functional principal components corresponding to 99 \% of the initial variability (variance) from the functional dataset. All the resulting values are summarized in Table 2.

\begin{table}[h]
\centering
\small
\setlength\tabcolsep{3pt}
\setlength\extrarowheight{2pt}
\captionsetup{format=sanslabel}

\begin{tabularx}{\textwidth}{ 
  >{\raggedright\arraybackslash}X 
  >{\raggedright\arraybackslash}X
  >{\raggedright\arraybackslash}X}
\midrule
\textsc{Parameters}   & \textsc{Screening}   & \textsc{Forecasting}    \\ \midrule
Initial length of time series      &  100 & 150 \\
\midrule
Number of knots in the basis decomposition   &  60 & 90\\

Number of functional principal components      & 10 & 15\\
Number of input values for the KNN    & 10 & 15\\

    \midrule          
\end{tabularx}
\caption{\textbf{Table 2} : Summary of the number of values taking action of the screening step (classification) and the forecasting step (regression) in the functional framework. The initial length of the time series can be compared with the number of inputs values of both KNNs, illustrating the dimensionality reduction from the functional data analysis.}
\end{table}

\subsection*{Validation metrics}
\paragraph{} All validation metrics within our study were obtained by tuning the hyperparameters of both KNN classification and regression \cite{Hyperparameters}. In practice, the number of nearest neighbors ($k$) is difficult to assess while initializing a random value. To deal with this, we used a cross validation (CV) through the \textit{GridsearchCV} function \cite{CV} available within the common \textit{Scikit-Learn} library of Python. We empirically set the number of CV splits at five to not increase the training time. Table 3 reflects the results obtained for the hyperparameters tuning. 

\begin{table}[h]
\centering
\small
\setlength\tabcolsep{3pt}
\setlength\extrarowheight{2pt}
\captionsetup{format=sanslabel}

\begin{tabularx}{\textwidth}{ 
  >{\raggedright\arraybackslash}X 
  >{\raggedright\arraybackslash}X
  >{\raggedright\arraybackslash}X}
\midrule
\textsc{Hyperparameter}   & \textsc{Screening}   & \textsc{Forecasting}    \\ \midrule
Number of nearest neighbors      &  $k$ = 28 & $k$ = 7 \\
Distance metric      & Manhattan & Minkowski\\
Power parameter      &  1 & 2\\
    \midrule          
\end{tabularx}
\caption{\textbf{Table 3} : Hyperparameters tuning from the cross validation, for the KNN classification and regression. Others KNN hyperparameters are set with initial values proposed by the function. }
\end{table}

\section*{Acknowledgements}
\paragraph{} The authors acknowledge the European Union's Horizon 2020 research and innovation programme for the funding support through the European Research Council (grant agreement 772873, “ARTISTIC” project : \href{https://www.erc-artistic.eu/}{ARTISTIC-ERC}). M.D. and A.A.F. acknowledge the ALISTORE European Research Institute for funding support. A.A.F. acknowledges the Institut Universitaire de France for the support. F.C. acknowledges the European Union's Horizon 2020 research, an innovation program under grant agreement No. 957189 (BIG MAP). We acknowledge Daphne Boursier at LRCS for the proofreading of the article and useful comments.

\section*{Author contributions statement}
\paragraph{} A.A.F. and M.D. had the original idea of the concept presented in this paper.  M.D. was in charge of the data driven development, the writing and data analysis. T.L. was in charge of the generation of NEMD simulations for NMC-based slurries and supported the writing and the data analysis. F.C. proofreaded the manuscript and provided useful comments implemented as inputs. J.X. and F.H. were in charge of the generation of NEMD simulations for LFP-based slurries. A.C.N. and H.O. supported the conceptualization of this work. A.A.F. obtained the funding (ERC Consolidator Grant), supervised the project and the preparation of the manuscript and carried out the revision.

\section*{Code availability}
\paragraph{} The codes and data are available for download from Github repositories upon publication of this paper : \\ \href{https://github.com/MarcDuquesnoy/FunctionalSimulations}{MarcDuquesnoy/FunctionalSimulations}, \href{https://github.com/ARTISTIC-ERC/FunctionalSimulations}{ARTISTIC-ERC/FunctionalSimulations}

\section*{Competing interests}
\paragraph{} The authors declare that they have no known competing financial interests or personal relationships that could have appeared to influence the work reported in this paper.

%
%
%
%
%

\bibliographystyle{unsrt}  
\bibliography{ms}

\end{document}


\begin{center}
    \huge \textbf{Functional Data-Driven Framework for Fast Forecasting of Electrode Slurry Rheology Simulated by Molecular Dynamics}
\end{center} 

\vspace{1.5cm}

\maketitle

{$^{1}$ \small Laboratoire de Réactivité et Chimie des Solides (LRCS), Université de Picardie Jules Verne UMR CNRS 7314, Hub de l’Energie, 80039 Amiens, France.}

{$^{2}$ \small Réseau sur le Stockage Electrochimique de l’Energie (RS2E), FR CNRS 3459, Hub de l’Energie, 80039 Amiens, France.}

{$^{3}$ \small Alistore-ERI European Research Institute, CNRS FR 3104, Hub de l’Energie, 80039 Amiens, France.}

{$^{4}$ \small Institut Universitaire de France, 103 Boulevard Saint Michel, 75005 Paris, France.}

{$^{*}$ \small Corresponding author : alejandro.franco@u-picardie.fr (Alejandro A. Franco)}

\hspace{10cm}
\\
\\
\\

\selectlanguage{english}

\newpage
\section{1. Methodology for the CGMD}
\paragraph{} The slurry structure was obtained after equilibrate the initial structure generated randomly in a simulation box. The initial structure was generated in order to match the experimental composition, in term of active material (\textit{i.e.} AM) and carbon binder domain (\textit{i.e.} CBD) mass ratio and AM particle size distribution (\textit{i.e.} PSD). The slurry equilibration was performed through a CGMD simulation in the NPT environment, the simulation box was relaxed until the desired temperature and pressure were reached (300 K and 1 atm), leading to the 3D slurry structure (Figure S1A)). Afterwards, a NEMD simulation was performed to our coarse-grained model to determine the viscosity at a given value of shear rate.  The deformation of the simulation box was applied according to the share rate (Figure S1B)) and the share strain was calculated during the deformation at each time step of the simulation as the ratio between the force acting on the upper part of the slurry on which the share rate was applied, divided by the surface of the slurry. The viscosity was calculated as the ratio between the share stress and the share rate at each time step of the simulation by using the nvt/sllod environment in LAMMPS. \\

\vspace{1cm}
\begin{figure}[!ht] 
	\captionsetup{format=sanslabel}
    \hbox to\hsize{\hss\includegraphics[width=16cm]{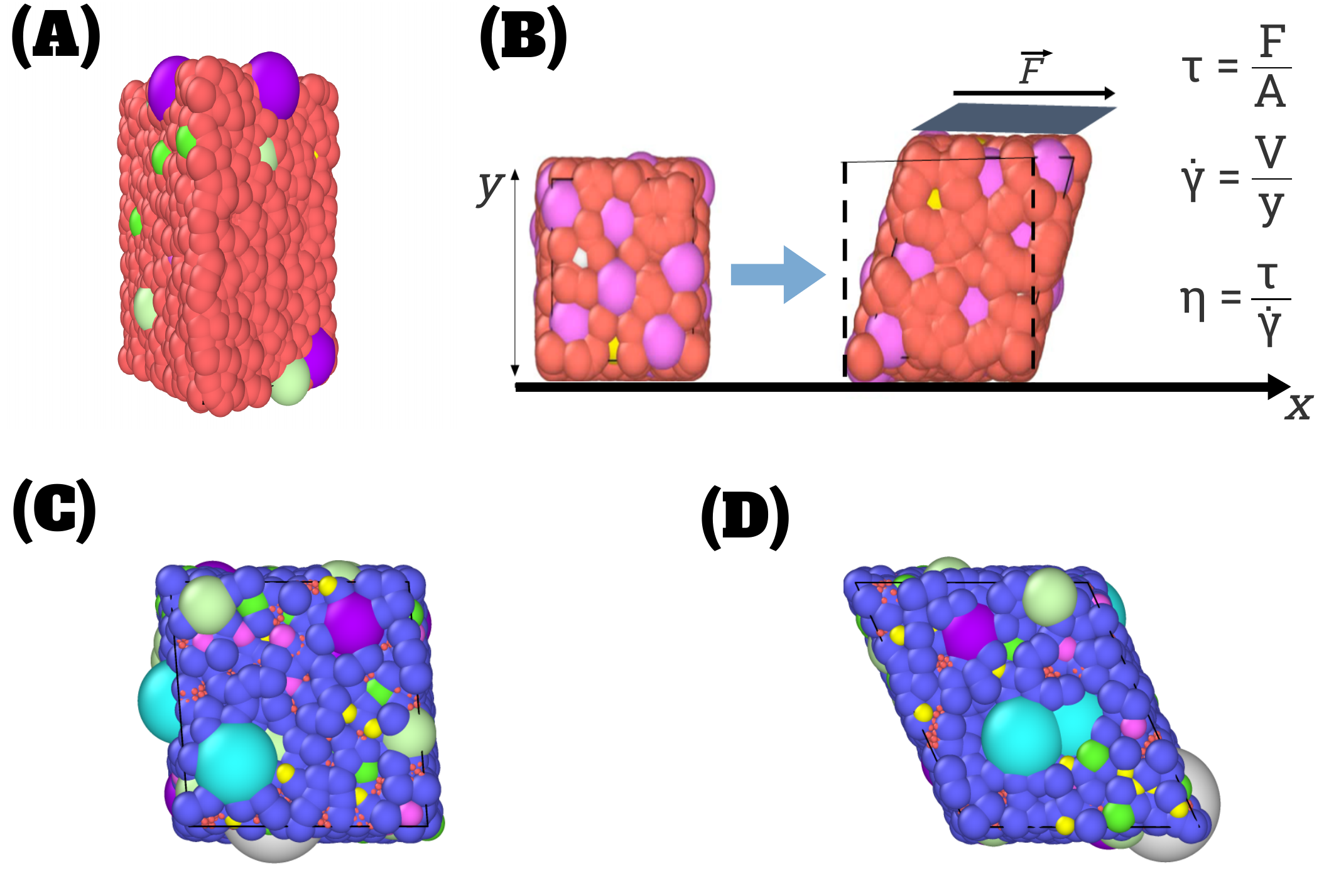}\hss} 
    \caption{ \textbf{Figure S1} : (A) Slurry 3D structure at 300 K ant 1 atm and (B) viscosity calculation. (C) Graphite slurry simulated for a shear rate equals to 26 $s^{-1}$. (D) Graphite slurry simulated for a shear rate equals to 163 $s^{-1}$.}
\end{figure}

\paragraph{} Figure S1C) and Figure S1D) show an example associated to a Graphite slurry for two different shear rates $\dot{\gamma}$. The deformation of the box is dependent of the chosen value.

\newpage
\section{2. K-Neareast-Neighbors for classification and regression}
\paragraph{} Each K-Nearest-Neighbors (KNN) model is trained by memorizing distances between training points in the variables space according to the chosen distance metric. To predict an output for new raw data that were not seen during the training step of the KNN model, the latter model uses the distances between training raw data from the $k$ nearest neighbors and assignes a majority vote (Figure S2A)) or an average value (Figure S2B)) respectively for classification and regression. \\

\begin{figure}[!ht] 
	\captionsetup{format=sanslabel}
    \hbox to\hsize{\hss\includegraphics[width=17cm, height=8.1cm]{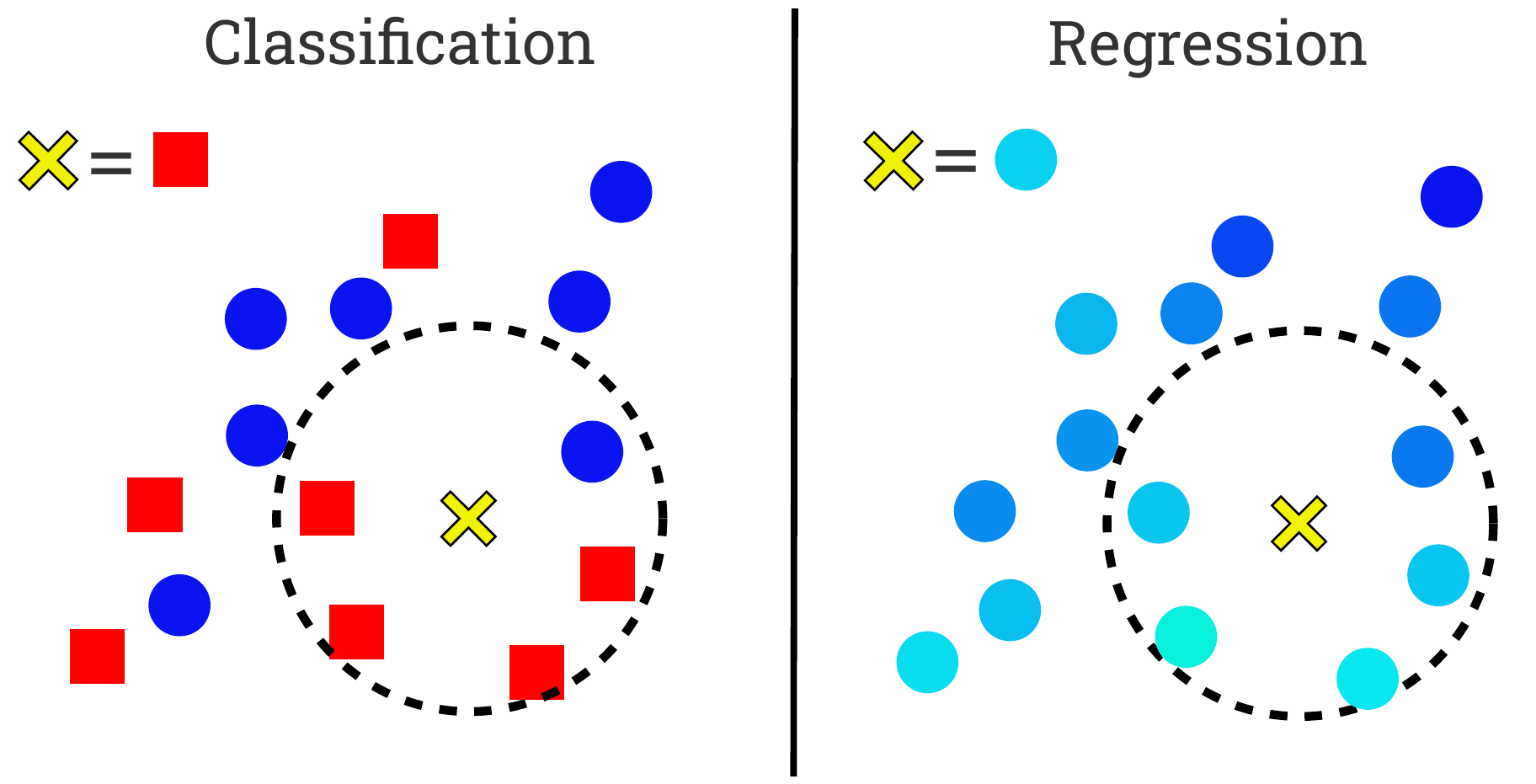}\hss} 
    \caption{ \textbf{Figure S2} : Example of the assessement of output from a KNN model in the case of classificiation and regression for a new data, color-code by a yellow cross. The number of nearest neighbors is set at $k=5$.}
\end{figure}

\bibliographystyle{unsrt}